\documentclass[12pt]{amsart}
\usepackage{amsmath,amssymb,esint,enumerate,graphicx,hyperref}
\usepackage[sort&compress,numbers]{natbib}

\hypersetup{pdftex,colorlinks=true,allcolors=blue}
\usepackage[all]{hypcap}

\usepackage[foot]{amsaddr}
\usepackage[sc]{mathpazo}

\newcommand\ack{\subsection*{Acknowledgment}}

\newcommand\BV{\boldsymbol} % Vector bold
\newcommand\Mach{\mathit{Ma}}
\newcommand\Rey{\mathit{Re}}
\newcommand\rhoHS{\rho_{H\!\!\;S}}
\newcommand\parderiv[2]{\frac{\partial #1}{\partial #2}}

\begin{document}

\author{Rafail V. Abramov}

\address{Department of Mathematics, Statistics and Computer Science,
University of Illinois at Chicago, 851 S. Morgan st., Chicago, IL 60607}

\email{abramov@uic.edu}

\title[Turbulence via intermolecular potential: Viscosity and the
  Reynolds number]{Turbulence via intermolecular
  potential:\\ Viscosity and transition range of the Reynolds number}

\begin{abstract}
Turbulence in fluids is an ubiquitous phenomenon, characterized by
spontaneous transition of a smooth, laminar flow to rapidly changing,
chaotic dynamics. In 1883, Reynolds experimentally demonstrated that,
in an initially laminar flow of water, turbulent motions emerge
without any measurable external disturbance. To this day, turbulence
remains a major unresolved phenomenon in fluid mechanics; in
particular, there is a lack of a mathematical model where turbulent
dynamics emerge naturally from a laminar flow. Recently, we proposed a
new theory of turbulence in gases, according to which turbulent
motions are created in an inertial gas flow by the mean field effect
of the intermolecular potential. In the current work, we investigate
the effect of viscosity in our turbulence model, by numerically
simulating the air flow at normal conditions in a straight pipe for
different values of the Reynolds number. We find that the transition
between the laminar and turbulent flows in our model occurs without
any deliberate perturbations as the Reynolds number increases from
2000 to 4000. As the simulated flow becomes turbulent, the decay rate
of the time averaged Fourier spectrum of the kinetic energy in our
model approaches Kolmogorov's inverse five-thirds law. Both results
are consistent with experiments and observations.
\end{abstract}

\maketitle

\section{Introduction}

Historically, the phenomenon of turbulence in fluids has been noted by
Leonardo da Vinci. However, its first scientifically documented
account appears to be due to \citet{Bou}, while the word
``turbulence'' itself has been suggested by \citet{Kel}. In his famous
experiment, \citet{Rey83} demonstrated that a laminar flow of water
spontaneously developed turbulent motions without any measurable
external disturbances, and that the onset of turbulence was reliably
associated with a sufficiently high value of the Reynolds number.
Later, Kolmogorov~\citep{Kol41b,Kol41c,Kol41a} found that the time-averaged
Fourier spectrum of the kinetic energy of turbulent air flow decayed
as the inverse five-thirds power of its wavenumber.

Despite overwhelming research efforts spanning multiple decades
\citep{Rey,Ric26,Tay,Tay38,KarHow38,Pra38,Obu41,Obu49,Cha49,Cor51,Kol62,Obu62,Kra66b,Kra66a,Saf67,Saf70,Man74},
the phenomenon of turbulence in gases and liquids remains unexplained;
namely, there does not exist an adequate fluid-mechanical model, for
either a liquid or a gas, where, at appropriate values of the Reynolds
number, turbulent flow naturally emerges from laminar initial and
boundary conditions in the absence of artificial external
disturbances. As examples, one can refer to recent works by
\citet{AviMoxLozAviBarHof,BarSonMukLemAviHof} and
\citet{KhaAnwHasSan}, where turbulent-like motions in a numerically
simulated flow had to be created artificially by deliberate
perturbations. In reality, turbulence emerges spontaneously by itself,
even if all reasonable measures were taken to preserve the laminarity
of the flow (e.g. the experiment of \citet{Rey83}) -- moreover, it was
precisely the spontaneity of the onset of turbulence which attracted
the world-wide scientific interest to this intriguing phenomenon.
Also, numerical simulations of artificially disturbed flows using
conventional equations of fluid mechanics fail to capture Kolmogorov's
power scaling of the kinetic energy spectra \citep{KhaAnwHasSan}.

In our recent works \citep{Abr22,Abr23,Abr24} we proposed a theory of
turbulence in a gas, where turbulent motions in an initially laminar,
inertial (that is, constant pressure) flow were created via the
average effect of gas molecules interacting by means of their
intermolecular potential $\phi(r)$. In our theory, this average effect
is expressed via the mean field potential $\bar\phi$ (which, of
course, depends on $\phi(r)$), whose gradient enters the equation for
the transport of momentum. According to our theory, in the absence of
the pressure gradient, the effect of $\bar\phi$ becomes the key
driving force of turbulent dynamics.  This novel effect is absent from
the conventional Euler and Navier--Stokes equations of fluid mechanics
because, in the Boltzmann--Grad limit \citep{Gra}, it is tacitly
assumed that the effect of $\bar\phi$ is negligible. However, we found
that, in our model of inertial gas flow, $\bar\phi$ produces turbulent
flow with Kolmogorov decay of the Fourier spectra of the kinetic
energy. Evidently, its effect is non-negligible and quantifiable.

Remarkably, \citet{Tsu} attempted to explain the creation of
turbulence via long-range correlations between molecules, but
Tsug\'e's result was restricted to an incompressible flow. On the
other hand, from what we discovered thus far, it appears that density
fluctuations are instrumental in the creation of turbulent dynamics.
In our past work \citep{Abr20}, we also considered long-range
interactions as a possible reason for the manifestation of turbulence.
However, we later found that even the hard sphere potential creates
turbulence in our model, which means that a typical intermolecular
potential is also capable of the same effect.

Thus far, our model of turbulent gas flow did not include viscous
effects. In the current work, we equip the momentum transport equation
of our model with the standard viscous term, which counteracts the
effect of the mean field potential (i.e. $\bar\phi$ creates turbulent
motions, while the viscosity dissipates them). We numerically simulate
the air flow at normal conditions within a straight pipe at different
values of the Reynolds number, and find that the transition between
the laminar and turbulent flow in our model occurs within the same
range of values of the Reynolds number as observed in
practice~\citep{Men,Let}.  Additionally, we find that, for turbulent
values of the Reynolds number, the rate of decay of the time-averaged
Fourier spectrum of the kinetic energy of the flow approaches the
famous Kolmogorov decay rate of the inverse five-thirds power of the
wavenumber.

The work is organized as follows. In Section~\ref{sec:inviscid_model},
we present the inviscid model of inertial flow with the mean field
potential, and demonstrate that, at a low Mach number, the effect of
the mean field potential in nondimensional variables is of the same
order as the rest of the terms. In Section~\ref{sec:viscous_model}, we
add viscosity into the momentum equation in a standard fashion. In
Section~\ref{sec:numerics}, we present the results of a numerical
simulation of the air flow at normal conditions in a straight pipe,
and show that the laminar-to-turbulent transition occurs as the
Reynolds number increases from 2000 to 4000.
Section~\ref{sec:summary} summarizes the results of this work.

\section{Our model of inertial turbulent gas flow}
\label{sec:inviscid_model}

In the context of our theory \citep{Abr22,Abr23,Abr24}, the turbulent flow
of a gas with density $\rho$ and velocity $\BV u$, at a constant
pressure $p_0$ (inertial flow) and in the absence of viscous effects,
is described by the following mass and momentum transport equations
\begin{equation}
\label{eq:density}
\parderiv\rho t+\nabla\cdot(\rho\BV u)=0,
\end{equation}
\begin{equation}
\label{eq:momentum}
\parderiv{(\rho\BV u)}t+\nabla\cdot(\rho\BV u^2)+\nabla\bar\phi=\BV 0.
\end{equation}
Above, observe that the momentum transport equation
\eqref{eq:momentum} possesses a novel term $\nabla\bar\phi$, which
replaces the pressure gradient in the conventional Euler or
Navier--Stokes equations. This term quantifies the average (or {\em
  mean field}) effect of the motion of molecules of mass $m$,
interacting via a potential $\phi(r)$. Therefore, in our preceding
works we referred to $\bar\phi$ as the {\em mean field potential}.

Assuming that the intermolecular potential $\phi(r)$ has the effective
range $\sigma$, we estimated the mean field potential $\bar\phi$ via
\begin{equation}
\label{eq:bphi}
\bar\phi=4p_0\rho/\rhoHS,
\end{equation}
where $\rhoHS=6 m/\pi \sigma^3$ is the density of the equivalent hard
sphere of mass $m$ and diameter~$\sigma$. Substituting $\bar\phi$ from
\eqref{eq:bphi} into \eqref{eq:momentum} yields
\begin{equation}
\label{eq:momentum_HS}
\parderiv{(\rho\BV u)}t+\nabla\cdot(\rho\BV u^2)+\frac{4p_0}
\rhoHS\nabla\rho=\BV 0.
\end{equation}
This novel effect is absent from the conventional Euler and
Navier--Stokes equations of fluid mechanics because, in the
Boltzmann--Grad hydrodynamic limit (see \citet{Gra}, pp.~352--353), it
is assumed that $\sigma^2/m\sim$ constant as $m$ and $\sigma$ are
taken to zero, and therefore, $\rhoHS\to\infty$. However, it can be
shown that, in the nondimensional variables, the effect of the mean
field potential in the inertial flow is comparable to the rest of the
terms at sufficiently low Mach numbers. In order to see this, we
rescale the variables in~\eqref{eq:density} and~\eqref{eq:momentum_HS}
in a standard fashion, by introducing the reference values of spatial
scale $L$, flow speed $U$, and density $\rho_0$:
\begin{equation}
\label{eq:rescalings}
\tilde t=Ut/L,\qquad\tilde{\BV x}=\BV x/L,\qquad\tilde\rho
=\rho/\rho_0,\qquad\tilde{\BV u}=\BV u/U.
\end{equation}
In addition, we denote the packing fraction $\eta$ and the Mach number
$\Mach$ via
\begin{equation}
\eta=\rho_0/\rhoHS,\qquad\Mach=U\sqrt{\rho_0/\gamma p_0},
\end{equation}
where $\gamma$ is the adiabatic exponent of the gas.  In the
nondimensional variables, the density equation \eqref{eq:density}
remains the same, while the momentum equation \eqref{eq:momentum_HS}
becomes
\begin{equation}
\parderiv{(\tilde\rho\tilde{\BV u})}{\tilde t}+\tilde\nabla\cdot(
\tilde\rho\tilde{\BV u}^2)+\alpha\tilde\nabla \tilde\rho=\BV 0,
\qquad\alpha=\frac{4\eta}{\gamma\Mach^2}.
\end{equation}
For air, $\gamma=1.4$, and $\rhoHS=1850$ kg/m$^3$ (see Eq.~(64) in our
work \citep{Abr23} for details). At normal conditions (sea level,
20$^\circ$C), we have $\rho_0=1.204$ kg/m$^3$, and $p_0=101.3$ kPa. As
a result, the packing fraction $\eta\approx 6.5\cdot 10^{-4}$. Also,
taking $U=30$ m/s, as in the numerical simulations below, we obtain
$\Mach\approx 8.74\cdot 10^{-2}$. Combining the estimates, we arrive
at $\alpha\approx 0.24$, for the settings of our numerical simulations
below. Moreover, reducing the speed of the flow to 15 m/s leads to
$\alpha\approx 1$. This confirms that the mean field effect of the
intermolecular potential at normal conditions and relatively low Mach
numbers is not negligible, and certainly does not vanish as presumed
in the Boltzmann--Grad limit.

\section{A model of inertial turbulent gas flow with viscosity}
\label{sec:viscous_model}

Due to the absence of viscosity, numerical solutions of
equations~(\ref{eq:density}) and~(\ref{eq:momentum_HS}) always develop
turbulent dynamics from an initially laminar flow
\citep{Abr22,Abr23,Abr24}.  In the current work, we add the standard
viscous term with the dynamic viscosity $\mu$ into the momentum
equation \eqref{eq:momentum_HS}:
\begin{equation}
\label{eq:momentum_HS_viscosity}
\parderiv{(\rho\BV u)}t+\nabla\cdot(\rho\BV u^2)+\frac{4p_0}\rhoHS
\nabla\rho=\nabla\cdot(\mu\nabla\BV u).
\end{equation}
According to the kinetic theory of gases \citep{HirCurBir}, $\mu$ is
proportional to the square root of the temperature. In a gas, the
product of temperature and density is proportional to the pressure,
which, in turn, is constant in an inertial gas flow. Therefore, in our
setting,
\begin{equation}
\mu=\mu_0\sqrt{\rho_0/\rho},
\end{equation}
where $\mu_0$ is the reference value of viscosity. In the
nondimensional variables, \eqref{eq:momentum_HS_viscosity} becomes
\begin{equation}
\label{eq:momentum_HS_viscosity_nondim}
\parderiv{(\tilde\rho\tilde{\BV u})}{\tilde t}+\tilde\nabla\cdot(
\tilde\rho\tilde{\BV u}^2)+\alpha\tilde\nabla \tilde\rho=\frac 1\Rey
\tilde\nabla\cdot\big(\tilde\rho^{-1/2}\tilde\nabla\tilde{\BV u}\big),
\end{equation}
where the Reynolds number $\Rey$ is given via
\begin{equation}
\label{eq:Rey}
\Rey=\rho_0UL/\mu_0.
\end{equation}
It is clear that, in this model, turbulent flow cannot emerge if the
coefficients of the forcing and dissipative terms are balanced, that
is, $\Rey=\alpha^{-1}\approx 4.1$ in our setting. However, for
$\Rey\gg\alpha^{-1}$, the manifestation of turbulent flow depends on
the geometry of the domain. In particular, it is known from
observations and practical engineering knowledge~\citep{Men}, that the
transition between laminar and turbulent gas flow in straight pipes
occurs within the range of values of the Reynolds number between 2000
and 4000, with the parameters $L$ and $U$ in~\eqref{eq:Rey} being the
width of the pipe, and the maximum speed of the flow, respectively.

We have to note that, in general, the presence of viscosity in a flow
induces pressure variations due to viscous friction; yet, above we
introduced viscosity into the momentum transport
equation~\eqref{eq:momentum_HS_viscosity} rather formally, without
accounting for such an effect. Thus, the system of transport equations
for the density~\eqref{eq:density} and
momentum~\eqref{eq:momentum_HS_viscosity} should not be viewed as a
practical method for accurate prediction of a real-world gas
flow. Instead, it should be treated as a ``proof-of-concept'' model,
whose purpose is to investigate how the relation between the mean
field potential forcing and viscous dissipation alone results in the
manifestation of turbulence in a gas flow, and at which values of the
Reynolds number such a transition occurs.

To confirm that our model is suitable for the above stated purpose,
first observe that, in an established laminar solution of the
conventional Navier--Stokes system, expressed in the nondimensional
variables, the term with the pressure gradient, induced by the viscous
friction, must be comparable to the viscous term itself (that is,
$\sim\Rey^{-1}$). In the numerical simulations below, we find that a
laminar flow still develops for $\Rey=1000$, which is also the
smallest value of the Reynolds number used. This, in turn, means that,
for $\alpha\approx 0.24$ in \eqref{eq:momentum_HS_viscosity_nondim},
the effect of the induced pressure gradient would be $\sim$ 240 times
smaller than that of the mean field potential term, and could,
therefore, be disregarded.

\section{Numerical simulations}
\label{sec:numerics}

To investigate whether or not our model predicts the transition to a
turbulent flow within a realistic range of values of the Reynolds
number, here we present numerical simulations of an inertial air flow
at normal conditions in a straight pipe. As in our recent works, we
use the appropriately modified \textit{rhoCentralFoam} solver
\citep{GreWelGasRee}, which uses the central discretization scheme of
\citet{KurTad}, with the flux limiter due to \citet{vanLee}. The
\textit{rhoCentralFoam} solver is a standard component of the
\textit{OpenFOAM} suite \citep{WelTabJasFur}.

\begin{figure}%
\includegraphics[width=\textwidth]{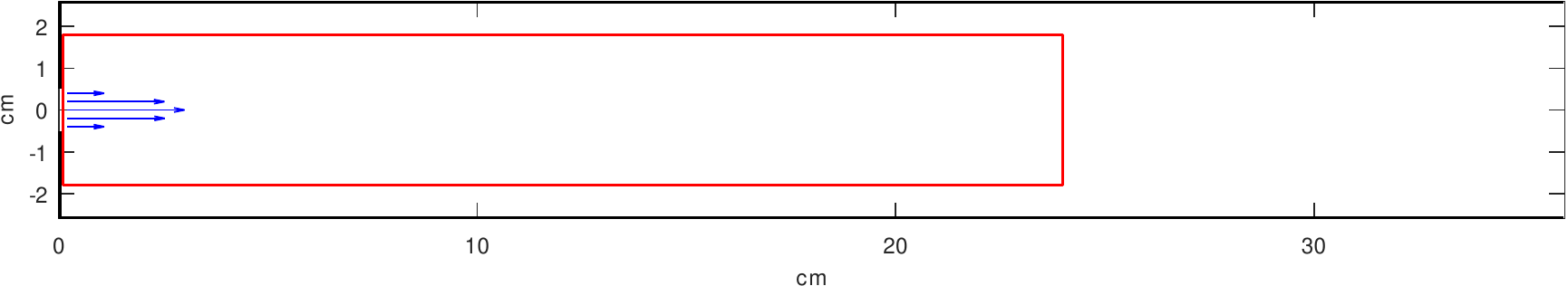}%
\caption{Longitudinal section of the computational domain. The domain
dimensions are $36\times 5.2\times 5.2$ cm$^3$. The inlet is on the
left, and the outlet is on the right. The pipe walls are shown via
thick black lines, so that both the inlet and outlet are visible.
The boundary of the Fourier spectrum measurement region is shown in
red. This region is a box of 24 cm in length, and $3.6\times 3.6$
cm$^2$ in cross-section.}%
\label{fig:domain}%
\end{figure}

Here, we simulate the inertial air flow in a straight pipe of a square
cross-section, using equations~\eqref{eq:density}
and~\eqref{eq:momentum_HS_viscosity}, with $p_0=101.3$ kPa, and
$\rhoHS=1850$ kg/m$^3$. The size of the pipe is $36\times 5.2\times
5.2$~cm$^3$. The domain is uniformly discretized in all directions
with the step of $0.8$ mm, which comprises $450\times 65\times
65=1,901,250$ finite volume cells in total.  The pipe is open-ended at
the outlet side, and has a wall at the inlet side, with the circular
inlet of 1~cm in diameter located in the middle of this wall. The
longitudinal section of the domain is shown in
Figure~\ref{fig:domain}.

The boundary conditions are the following. The density is set to
$\rho=1.204$ kg/m$^3$ at the outlet, and has zero normal derivative at
the inlet and the walls. The velocity has zero normal derivative at
the outlet, no-slip condition at the walls, and a radially symmetric
parabolic profile at the inlet, with the maximum of 30 m/s in the
middle of the inlet, directed along the axis of the pipe. The diameter
of the inlet, and the speed of the entering flow correspond to the
experiment by \citet{BucVel}. Initially, the gas inside the pipe is at
rest, with zero velocity and uniform density set to 1.204 kg/m$^3$.

We conducted numerical simulations of~\eqref{eq:density}
and~\eqref{eq:momentum_HS_viscosity} for the values of the Reynolds
number $\Rey=1000$, $2000$, $3000$ and $4000$, by setting the
reference viscosity $\mu_0$ to $1.872\cdot 10^{-3}$, $9.36\cdot
10^{-4}$, $6.24\cdot 10^{-4}$ and $4.68\cdot 10^{-4}$ kg/m s,
respectively. We integrated equations~\eqref{eq:density}
and~\eqref{eq:momentum_HS_viscosity} forward in time using the
explicit (forward Euler) scheme for the advection and mean field
potential forcing terms, and implicit (backward Euler) scheme for the
viscous term. The forward time stepping of the scheme was adaptive,
with the time step set to 20\% of the maximal allowed by the Courant
number.

Here, observe that the goal of the simulation is not the local
accuracy of the solution, but rather the accurate capture of the
statistical regime of the dynamics (and, in particular, the transition
to nonlinear chaos). This means that the numerical integration scheme
should be chosen so as to avoid the introduction of artificial damping
into the advection part of the system. Due to this reason, the simple
forward Euler method appears to be better for such a specific purpose
than more advanced numerical integration schemes, such as the 4th
order Runge--Kutta method, since the latter tend to introduce
artificial damping into numerical solutions as a result of their
superior stability properties \citep{YuTsaHsi}.

\subsection{Results}

\begin{figure}%
\includegraphics[width=\textwidth]{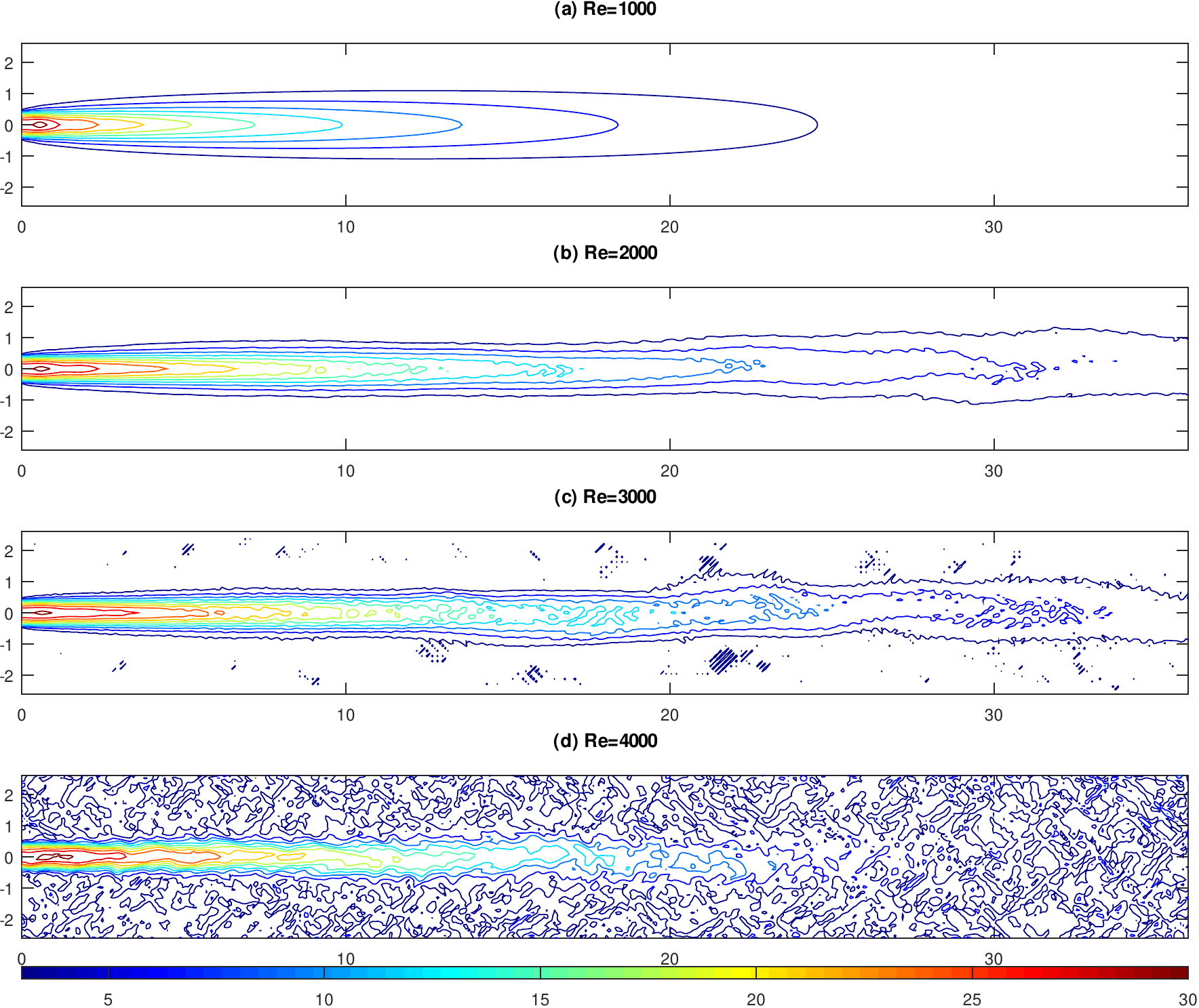}%
\caption{Speed of the flow (m/s), expressed in the form of level
  curves, and captured in the longitudinal symmetry plane of the pipe
  at the elapsed time $t=0.15$~s for (a) $\Rey=1000$, (b) $\Rey=2000$,
  (c) $\Rey=3000$ and (d) $\Rey=4000$.}%
\label{fig:speed}
\end{figure}

We found that the flow fully developed by the elapsed time $t=0.07$
seconds in all simulations. In Figure~\ref{fig:speed}, we show four
snapshots of the speed of the flow, taken in the longitudinal symmetry
plane of the pipe at $t=0.15$ s, which illustrate the transition
between the laminar and turbulent flow. For $\Rey=1000$, the flow is
laminar, as evidenced by smoothness of the level curves, and symmetric
relative to the axis of the pipe. For $\Rey=2000$, the symmetry of the
flow is broken, and small intermittent fluctuations appear in the
otherwise laminar flow. These fluctuations become larger and more
numerous for $\Rey=3000$; for $\Rey=4000$, the flow is fully
turbulent. Clearly, the range of values of the Reynolds number, at
which the turbulent transition occurs in our model, agrees with
observations \citep{Men}. The breaking of the flow symmetry during the
transition to turbulence is likely associated with the onset of chaos
in the dynamics, and, in numerical simulations, happens due to
exponentially growing machine round-off errors. The hypothesis that
turbulence is a manifestation of nonlinear chaos has also been
discussed in the literature (see \citet{Let} and references therein).

\begin{figure}
\includegraphics[width=\textwidth]{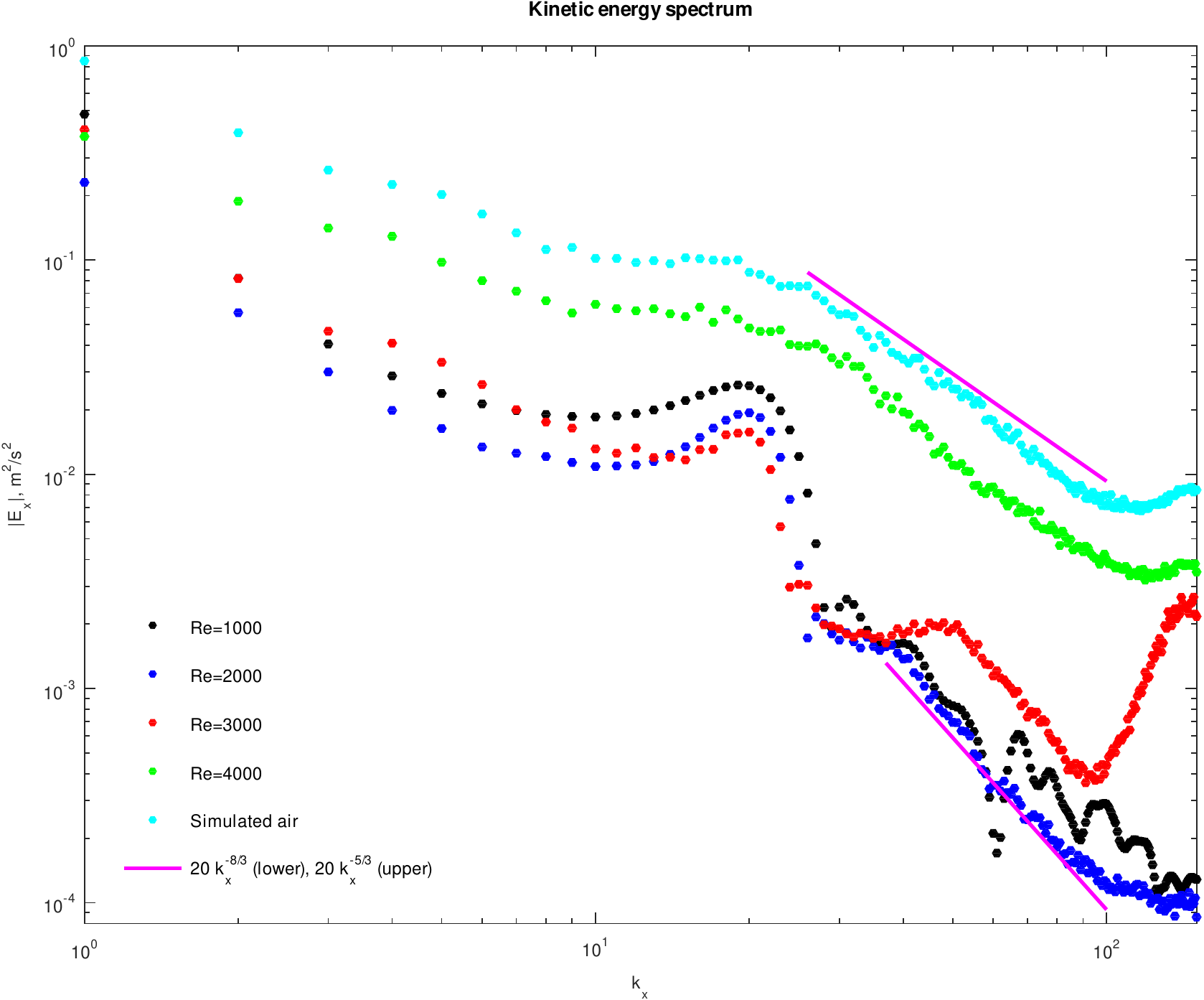}%
\caption{The Fourier spectra of the kinetic energy, averaged between
  $0.1$ and $0.2$ s of the elapsed time, for $\Rey=1000$, $2000$,
  $3000$ and $4000$, as well as air ($\mu_0=1.825\cdot 10^{-5}$ kg/m
  s). The power decay slopes $E_0 k_x^{-5/3}$ and $E_0 k_x^{-8/3}$,
  with $E_0=20$ m$^2$/s$^2$, are added for reference.}%
\label{fig:energy}
\end{figure}%

In addition to the snapshots of the speed of the flow, we computed the
time averages of its kinetic energy spectrum. The computation was done
within the central core of the pipe of $3.6\times 3.6$ cm$^2$ in
cross-section, extending between 0 and 24 cm of the length of the pipe
(shown in red in Figure~\ref{fig:domain}), and thus largely containing
the jet stream.  The time averaging was carried out in the interval
between $0.1$ and $0.2$ seconds of the elapsed time. For the detailed
description of the energy spectrum computation, see our recent works
\citep{Abr22,Abr23,Abr24}.

In Figure~\ref{fig:energy}, we show time averages of the computed
Fourier spectra of the streamwise component of the kinetic energy for
the same simulated flows, which are displayed as functions of their
Fourier wavenumber $k_x$ in the longitudinal direction of the pipe on
a logarithmic scale. In addition, we show the kinetic energy spectrum
for the simulation with $\mu_0=1.825\cdot 10^{-5}$ kg/m s ($\Rey\sim
10^5$), which corresponds to the viscosity of air at normal
conditions. For the reference, we show two power slope lines, given
via $E_0 k_x^{-8/3}$ and $E_0 k_x^{-5/3}$, which share the same
empirically chosen scaling constant $E_0=20$ m$^2$/s$^2$.

At the large scale Fourier wavenumbers, the structures of the kinetic
energy spectra of all computed flows are similar, while the major
differences are observed at moderate and small scales.  For
$\Rey=1000$, $2000$ and $3000$, the kinetic energy spectrum at small
scales generally appears to match the $k^{-8/3}$-decay slope, with the
following variations. In the $\Rey=2000$ regime, the spectrum decays
along the $k^{-8/3}$-slope rather monotonously, whereas in the
$\Rey=1000$ regime (which is fully laminar) the spectrum also exhibits
oscillations around this slope. As we hypothesized in \citep{Abr24},
the latter could be a manifestation of the quasi-periodic dynamics at
the unstable Fourier wavenumbers, with the periodicity of orbits
destroyed by chaos as $\Rey$ increases to 2000. In the regime with
$\Rey=3000$, an unusual growth of the energy spectrum is observed at
small scales. Remarkably, the rate of decay of the kinetic energy
spectrum for the turbulent regime $\Rey=4000$, as well as that of the
air, approach Kolmogorov's $k^{-5/3}$-slope.

\section{Summary}
\label{sec:summary}

In the current work, we formally introduce viscosity into our model of
turbulence via an intermolecular potential \citep{Abr22,Abr23,Abr24},
to investigate the transition between laminar and turbulent air flows
with varying Reynolds number. We numerically simulate the air flow at
normal conditions in a straight pipe at different values of the
Reynolds number, and find that the transition into turbulent flow
occurs when the Reynolds number increases from 2000 to 4000. This
appears to be consistent with observations, experiments and practical
knowledge. Additionally, we find that, in our model, the corresponding
rate of decay of the time-averaged Fourier transform of the streamwise
kinetic energy at small scales changes from $k^{-8/3}$-slope towards
$k^{-5/3}$-slope (Kolmogorov's law), as the flow transitions from
laminar to turbulent.

The results of our work seem to be encouraging. For the first time in
history, we created a model of compressible gas flow in the form of
fluid mechanics equations, where, first, turbulent dynamics emerge
from an initially laminar flow at appropriate values of the Reynolds
number naturally and without the help of artificial disturbances, and,
second, the rate of decay of the Fourier spectrum of the kinetic
energy of turbulent flow in our model matches that observed in nature.

At the same time, our model has its limitations, as it describes the
inertial gas flow; in a realistic flow, the pressure generally
fluctuates, even if slightly. However, other models of fluid mechanics
have their own limitations -- for example, both the incompressible and
compressible Euler equations are incompatible with the process of
convection (the density is constant in the former, and increases when
the air warms up in the latter). Yet, such models are widely used,
because they describe specific features of the flow which are needed
for relevant practical applications. Similarly, our model may find
its own use -- perhaps, as a ``stepping stone'', improving our general
understanding of the fluid mechanics of turbulence.

\ack This work was supported by the Simons Foundation grant \#636144.


\begin{thebibliography}{41}
\providecommand{\natexlab}[1]{#1}
\providecommand{\url}[1]{\texttt{#1}}
\expandafter\ifx\csname urlstyle\endcsname\relax
  \providecommand{\doi}[1]{doi: #1}\else
  \providecommand{\doi}{doi: \begingroup \urlstyle{rm}\Url}\fi

\bibitem[Boussinesq(1877)]{Bou}
J.~Boussinesq.
\newblock Essai sur la th\'eorie des eaux courantes.
\newblock \emph{M\'emoires pr\'esent\'es par divers savants \`a l'Acad\'emie
  des Sciences}, XXIII\penalty0 (1):\penalty0 1--680, 1877.

\bibitem[Thomson(1887)]{Kel}
W.~Thomson.
\newblock {XLV}. {O}n the propagation of laminar motion through a turbulently
  moving inviscid liquid.
\newblock \emph{Philos. Mag. Series 5}, 24\penalty0 (149):\penalty0 342--353,
  1887.

\bibitem[Reynolds(1883)]{Rey83}
O.~Reynolds.
\newblock An experimental investigation of the circumstances which determine
  whether the motion of water shall be direct or sinuous, and of the law of
  resistance in parallel channels.
\newblock \emph{Proc. R. Soc. Lond.}, 35\penalty0 (224--226):\penalty0 84--99,
  1883.

\bibitem[Kolmogorov(1941{\natexlab{a}})]{Kol41b}
A.N. Kolmogorov.
\newblock Decay of isotropic turbulence in an incompressible viscous fluid.
\newblock \emph{Dokl. Akad. Nauk SSSR}, 31:\penalty0 538--541,
  1941{\natexlab{a}}.

\bibitem[Kolmogorov(1941{\natexlab{b}})]{Kol41c}
A.N. Kolmogorov.
\newblock Energy dissipation in locally isotropic turbulence.
\newblock \emph{Dokl. Akad. Nauk SSSR}, 32:\penalty0 19--21,
  1941{\natexlab{b}}.

\bibitem[Kolmogorov(1941{\natexlab{c}})]{Kol41a}
A.N. Kolmogorov.
\newblock Local structure of turbulence in an incompressible fluid at very high
  {R}eynolds numbers.
\newblock \emph{Dokl. Akad. Nauk SSSR}, 30:\penalty0 299--303,
  1941{\natexlab{c}}.

\bibitem[Reynolds(1895)]{Rey}
O.~Reynolds.
\newblock On the dynamical theory of incompressible viscous fluids and the
  determination of the criterion.
\newblock \emph{Phil. Trans. Roy. Soc. A}, 186:\penalty0 123--164, 1895.

\bibitem[Richardson(1926)]{Ric26}
L.F. Richardson.
\newblock Atmospheric diffusion shown on a distance-neighbour graph.
\newblock \emph{Proc. Roy. Soc. London A}, 110:\penalty0 709--737, 1926.

\bibitem[Taylor(1935)]{Tay}
G.I. Taylor.
\newblock Statistical theory of turbulence.
\newblock \emph{Proc. Roy. Soc. London A}, 151\penalty0 (873):\penalty0
  421--444, 1935.

\bibitem[Taylor(1938)]{Tay38}
G.I. Taylor.
\newblock The spectrum of turbulence.
\newblock \emph{Proc. Roy. Soc. London A}, 164:\penalty0 476--490, 1938.

\bibitem[von K\'arm\'an and Howarth(1938)]{KarHow38}
T.~von K\'arm\'an and L.~Howarth.
\newblock On the statistical theory of isotropic turbulence.
\newblock \emph{Proc. Roy. Soc. London A}, 164:\penalty0 192--215, 1938.

\bibitem[Prandtl(1938)]{Pra38}
L.~Prandtl.
\newblock Beitrag zum {T}urbulenzsymposium.
\newblock In J.P. Den~Hartog and H.~Peters, editors, \emph{Proceedings of the
  Fifth International Congress on Applied Mechanics}, pages 340--346, Cambridge
  MA, 1938. John Wiley, New York.

\bibitem[Obukhov(1941)]{Obu41}
A.M. Obukhov.
\newblock On the distribution of energy in the spectrum of a turbulent flow.
\newblock \emph{Izv. Akad. Nauk SSSR Ser. Geogr. Geofiz.}, 5:\penalty0
  453--466, 1941.

\bibitem[Obukhov(1949)]{Obu49}
A.M. Obukhov.
\newblock Structure of the temperature field in turbulent flow.
\newblock \emph{Izv. Akad. Nauk SSSR Ser. Geogr. Geofiz.}, 13:\penalty0 58--69,
  1949.

\bibitem[Chandrasekhar(1949)]{Cha49}
S.~Chandrasekhar.
\newblock On {H}eisenberg's elementary theory of turbulence.
\newblock \emph{Proc. Roy. Soc.}, 200:\penalty0 20--33, 1949.

\bibitem[Corrsin(1951)]{Cor51}
S.~Corrsin.
\newblock On the spectrum of isotropic temperature fluctuations in an isotropic
  turbulence.
\newblock \emph{J. Appl. Phys.}, 22\penalty0 (4):\penalty0 469--473, 1951.

\bibitem[Kolmogorov(1962)]{Kol62}
A.N. Kolmogorov.
\newblock A refinement of previous hypotheses concerning the local structure of
  turbulence in a viscous incompressible fluid at high {R}eynolds number.
\newblock \emph{J. Fluid Mech.}, 13\penalty0 (1):\penalty0 82--85, 1962.

\bibitem[Obukhov(1962)]{Obu62}
A.M. Obukhov.
\newblock Some specific features of atmospheric turbulence.
\newblock \emph{J. Geophys. Res.}, 67\penalty0 (8):\penalty0 3011--3014, 1962.

\bibitem[Kraichnan(1966{\natexlab{a}})]{Kra66b}
R.H. Kraichnan.
\newblock Dispersion of particle pairs in homogeneous turbulence.
\newblock \emph{Phys. Fluids}, 9:\penalty0 1937--1943, 1966{\natexlab{a}}.

\bibitem[Kraichnan(1966{\natexlab{b}})]{Kra66a}
R.H. Kraichnan.
\newblock Isotropic turbulence and inertial range structure.
\newblock \emph{Phys. Fluids}, 9:\penalty0 1728--1752, 1966{\natexlab{b}}.

\bibitem[Saffman(1967)]{Saf67}
P.G. Saffman.
\newblock The large-scale structure of homogeneous turbulence.
\newblock \emph{J. Fluid Mech.}, 27\penalty0 (3):\penalty0 581--593, 1967.

\bibitem[Saffman(1970)]{Saf70}
P.G. Saffman.
\newblock A model for inhomogeneous turbulent flow.
\newblock \emph{Proc. Roy. Soc. London A}, 317:\penalty0 417--433, 1970.

\bibitem[Mandelbrot(1974)]{Man74}
B.B. Mandelbrot.
\newblock Intermittent turbulence in self-similar cascades; divergence of high
  moments and dimension of the carrier.
\newblock \emph{J. Fluid Mech.}, 62\penalty0 (2):\penalty0 331--358, 1974.

\bibitem[Avila et~al.(2011)Avila, Moxey, de~Lozar, Avila, Barkley, and
  Hof]{AviMoxLozAviBarHof}
K.~Avila, D.~Moxey, A.~de~Lozar, M.~Avila, D.~Barkley, and B.~Hof.
\newblock The onset of turbulence in pipe flow.
\newblock \emph{Science}, 333:\penalty0 192--196, 2011.

\bibitem[Barkley et~al.(2015)Barkley, Song, Mukund, Lemoult, Avila, and
  Hof]{BarSonMukLemAviHof}
D.~Barkley, B.~Song, V.~Mukund, G.~Lemoult, M.~Avila, and B.~Hof.
\newblock The rise of fully turbulent flow.
\newblock \emph{Nature}, 526:\penalty0 550--553, 2015.

\bibitem[Khan et~al.(2021)Khan, Anwer, Hasan, and Sanghi]{KhaAnwHasSan}
H.H. Khan, S.F. Anwer, N.~Hasan, and S.~Sanghi.
\newblock Laminar to turbulent transition in a finite length square duct
  subjected to inlet disturbance.
\newblock \emph{Phys. Fluids}, 33:\penalty0 065128, 2021.

\bibitem[Abramov(2021{\natexlab{a}})]{Abr22}
R.V. Abramov.
\newblock Macroscopic turbulent flow via hard sphere potential.
\newblock \emph{AIP Adv.}, 11\penalty0 (8):\penalty0 085210,
  2021{\natexlab{a}}.

\bibitem[Abramov(2021{\natexlab{b}})]{Abr23}
R.V. Abramov.
\newblock Turbulence in large-scale two-dimensional balanced hard sphere gas
  flow.
\newblock \emph{Atmosphere}, 12\penalty0 (11):\penalty0 1520,
  2021{\natexlab{b}}.

\bibitem[Abramov(2022)]{Abr24}
R.V. Abramov.
\newblock Creation of turbulence in polyatomic gas flow via an intermolecular
  potential.
\newblock Preprint: \url{https://arxiv.org/abs/2201.07175}, 2022.

\bibitem[Grad(1949)]{Gra}
H.~Grad.
\newblock On the kinetic theory of rarefied gases.
\newblock \emph{Comm. Pure Appl. Math.}, 2\penalty0 (4):\penalty0 331--407,
  1949.

\bibitem[Tsug\'e(1974)]{Tsu}
S.~Tsug\'e.
\newblock Approach to the origin of turbulence on the basis of two-point
  kinetic theory.
\newblock \emph{Phys. Fluids}, 17\penalty0 (1):\penalty0 22--33, 1974.

\bibitem[Abramov(2020)]{Abr20}
R.V. Abramov.
\newblock Turbulent energy spectrum via an interaction potential.
\newblock \emph{J. Nonlinear Sci.}, 30:\penalty0 3057--3087, 2020.

\bibitem[Menon(2005)]{Men}
E.S. Menon.
\newblock \emph{Gas Pipeline Hydraulics}.
\newblock Taylor \& Francis, Boca Raton, FL, 2005.

\bibitem[Letellier(2017)]{Let}
C.~Letellier.
\newblock Intermittency as a transition to turbulence in pipes: A long
  tradition from {R}eynolds to the 21st century.
\newblock \emph{C.R. Mecanique}, 345:\penalty0 642--659, 2017.

\bibitem[Hirschfelder et~al.(1964)Hirschfelder, Curtiss, and Bird]{HirCurBir}
J.O. Hirschfelder, C.F. Curtiss, and R.B. Bird.
\newblock \emph{The Molecular Theory of Gases and Liquids}.
\newblock Wiley, 1964.

\bibitem[Greenshields et~al.(2010)Greenshields, Weller, Gasparini, and
  Reese]{GreWelGasRee}
C.J. Greenshields, H.G. Weller, L.~Gasparini, and J.M. Reese.
\newblock Implementation of semi-discrete, non-staggered central schemes in a
  colocated, polyhedral, finite volume framework, for high-speed viscous flows.
\newblock \emph{Int. J. Numer. Methods Fluids}, 63\penalty0 (1):\penalty0
  1--21, 2010.

\bibitem[Kurganov and Tadmor(2001)]{KurTad}
A.~Kurganov and E.~Tadmor.
\newblock New high-resolution central schemes for nonlinear conservation laws
  and convection--diffusion equations.
\newblock \emph{J. Comput. Phys.}, 160:\penalty0 241--282, 2001.

\bibitem[van Leer(1974)]{vanLee}
B.~van Leer.
\newblock Towards the ultimate conservative difference scheme, {II}:
  {M}onotonicity and conservation combined in a second order scheme.
\newblock \emph{J. Comput. Phys.}, 17:\penalty0 361--370, 1974.

\bibitem[Weller et~al.(1998)Weller, Tabor, Jasak, and Fureby]{WelTabJasFur}
H.G. Weller, G.~Tabor, H.~Jasak, and C.~Fureby.
\newblock A tensorial approach to computational continuum mechanics using
  object-oriented techniques.
\newblock \emph{Computers in Physics}, 12\penalty0 (6):\penalty0 620--631,
  1998.

\bibitem[Buchhave and Velte(2017)]{BucVel}
P.~Buchhave and C.M. Velte.
\newblock Measurement of turbulent spatial structure and kinetic energy
  spectrum by exact temporal-to-spatial mapping.
\newblock \emph{Phys. Fluids}, 29\penalty0 (8):\penalty0 085109, 2017.

\bibitem[Yu et~al.(1992)Yu, Tsai, and Hsieh]{YuTsaHsi}
S.-T. Yu, Y.-L.P. Tsai, and K.C. Hsieh.
\newblock Runge--{K}utta methods combined with compact difference schemes for
  the unsteady {E}uler equations.
\newblock In \emph{28th Joint Propulsion Conference and Exhibit}, pages 1--28.
  AIAA, SAE, ASME, and ASEE, 1992.

\end{thebibliography}
\end{document}